# Temperature-dependent structural property and power factor of n type thermoelectric $Bi_{0.90}Sb_{0.10}$ and $Bi_{0.86}Sb_{0.14}$ alloys

K. Malik[1], Diptasikha Das[1], S. Bandyopadhyay[1,2], P. Mandal[3], A. K. Deb[4], Velaga Srihari[3] and Aritra Banerjee[1,2,a)]

[1]Department of Physics, University of Calcutta, 92 A P C Road, Kolkata 700 009, India

[2]Center for Research in Nanoscience and Nanotechnology, University of Calcutta, JD-2, Sector-III, Saltlake City, Kolkata 700 098, India

[3]Saha Institute of Nuclear Physics, 1/AF Bidhannagar, Kolkata 700 064, India

[4]Department of Physics, Raiganj College (University College), Uttar Dinajpur 733 134, India

## ABSTRACT

Thermal variation of structural property, linear thermal expansion coefficient ($\alpha$), resistivity ($\rho$), thermopower (S) and power factor (PF) of polycrystalline $Bi_{1-x}Sb_x$ (x=0.10, 0.14) samples are reported. Temperature-dependent powder diffraction experiments indicate that samples do not undergo any structural phase transition. Rietveld refinement technique has been used to perform detailed structural analysis. Temperature dependence of $\alpha$ is found to be stronger for $Bi_{0.90}Sb_{0.10}$. Also, PF for direct band gap $Bi_{0.90}Sb_{0.10}$ is higher as compared to that for indirect band gap $Bi_{0.86}Sb_{0.14}$. Role of electron-electron and electron-phonon scattering on $\rho$, S, and PF have been discussed.

[a)] Author to whom correspondence should be addressed. Electronic mail: arbphy@caluniv.ac.in

Thermoelectric (TE) devices are potential candidates for solid state cooling and power generation from waste heat. They generate electric voltage from the temperature differences and vice versa. These devices use two types of semiconductor "legs", n type and p type, connected in series. The performance of a TE material is evaluated by the figure of merit $ZT = S^2/\rho\kappa$, where S, ρ, and κ are the Seebeck coefficient, electrical resistivity, and thermal conductivity of the semiconducting material, respectively.[1] An efficient TE material should posses high ZT. Worldwide there is a resurgence in activity for revealing the underlying physics towards obtaining higher ZT.[2] By maximizing the power-factor $PF= S^2/\rho$ and/or lowering the thermal conductivity, efficiency of a TE device can be improved.[3,4] Different scattering mechanisms such as electron-electron (*e-e*) and electron-phonon (*e-ph*) directly influence the Seebeck coefficient, resistivity, and thermal conductivity[2] of a material and thus should be intimately related to the enhancement of PF and ZT. Among semimetals and narrow gap semiconductors, $Bi_{1-x}Sb_x$ alloys are well known n-type TE material. Their unique properties such as small band gap, high mobility, small effective mass and reduced thermal conductivity made them attracting as one of the best n type thermoelectric materials around 200 K.[5] $Bi_{1-x}Sb_x$ posses an interesting band structure and can be either a semimetal or semiconductor depending on Sb concentration.[6] Precisely, semiconducting $Bi_{1-x}Sb_x$ ($0.08 \leq x \leq 0.22$) alloys with 8 to 12 at.% Sb content are a direct band gap semiconductor but for higher Sb concentration with $0.14 \leq x \leq 0.22$, they become an indirect gap semiconductor[6,7].

Although there are several efforts in estimating PF and ZT for $Bi_{1-x}Sb_x$ alloys, not much effort has been given in evaluating the effect of different scattering mechanisms on PF and hence on ZT. In this letter, we demonstrate the role of *e-e* and *e-ph* scatterings on PF for two typical $Bi_{1-x}Sb_x$ alloys, *viz.*, the direct band gap $Bi_{0.90}Sb_{0.10}$ and the indirect band gap $Bi_{0.86}Sb_{0.14}$. This is very important for designing an efficient TE device. PF has been estimated from the temperature dependence of resistivity and thermopower. The *e-e* and *e-ph* scattering terms are extracted by fitting the ρ(T) and S(T) data. In addition, thermal expansion, the measure of modification of shape and size with temperature variation, is one of the important parameters in device fabrication. Hence, in this letter, we also present temperature dependence of lattice parameters and estimate the values of linear thermal expansion coefficient (α) for the said $Bi_{1-x}Sb_x$ alloys.

Polycrystalline $Bi_{0.90}Sb_{0.10}$ and $Bi_{0.86}Sb_{0.14}$ samples have been synthesized by solid-state reaction method.[6] Low-temperature powder diffraction experiments are carried out using synchrotron X-ray diffraction (XRD) facility at Indian beamline BL-18B, Photon Factory, Japan. The sample was illuminated by a monochromatic X-ray beam with wavelength λ=0.992 Å. All the XRD measurements have been performed in $10^0 \leq 2\theta \leq 70^0$ range. The data for powder diffraction experiments have been collected during heating cycle. The Rietveld method (MAUD software) has been used to refine the structural parameters. The instrumental profile is estimated using high purity Si powder as an internal standard.[8] Resistivity has been measured by conventional four-probe technique. Thermopower measurements are performed by differential technique, where a small temperature gradient (ΔT) is created across the sample and the voltage (ΔE) developed between the hot and cold end is measured. At a particular temperature, the thermopower is calculated from the slope of ΔE versus ΔT plot to eliminate any contribution from spurious emf.

For both the samples, powder diffraction experiments are carried out at three different temperatures viz., 300, 130 and 15 K and the corresponding XRD patterns are presented in Fig. 1. So far very limited effort has been given to evaluate the structural properties of $Bi_{1-x}Sb_x$ using Rietveld refinement technique.[6] Though, there are some reports on the temperature dependence of structural properties of several well known thermoelectric materials viz., $Bi_2Te_3$, $Sb_2Te_3$, and $Bi_2Se_3$[9-11], all the reports on structural properties of single crystals,[5] thin films[12] and polycrystalline samples[6,13] of $Bi_{1-x}Sb_x$ are at room temperature. In order to study the effect of temperature on the structural parameters of $Bi_{1-x}Sb_x$, in depth analysis using Rietveld refinement has been performed. At room temperature, $Bi_{1-x}Sb_x$ crystallize in rhombohedral $A_7$ type structure.[6,12] In our maiden attempt, the powder diffraction experiments reveal that the studied $Bi_{1-x}Sb_x$ alloys retain their rhombhedral $A_7$ type structure down to the lowest temperature measured, indicating that the $Bi_{1-x}Sb_x$ samples do not undergo any structural phase transition in the temperature range 15-300 K. Experimental XRD curve for different temperatures and the theoretical curve obtained after refinement are shown in Fig.S1 and Fig.S2 (see the supplementary information for Figure S1 and Figure S2).[14] The refinement has been done using both atomic position and substitution. Space group $R\bar{3}m$ and point group $D_{3d}$ with hexagonal coordinate system are used for refinement.[5] The obtained reliability parameters ($R_w$, $R_b$, $R_{exp}$) and the goodness of fit (GoF) or $\chi^2$ reflect high quality structural refinement [Figs. S1, S2].[14] As expected, the lattice parameters $a_H$ and $c_H$, and hence the corresponding unit cell volume (V) decrease with decreasing temperature [Table I].

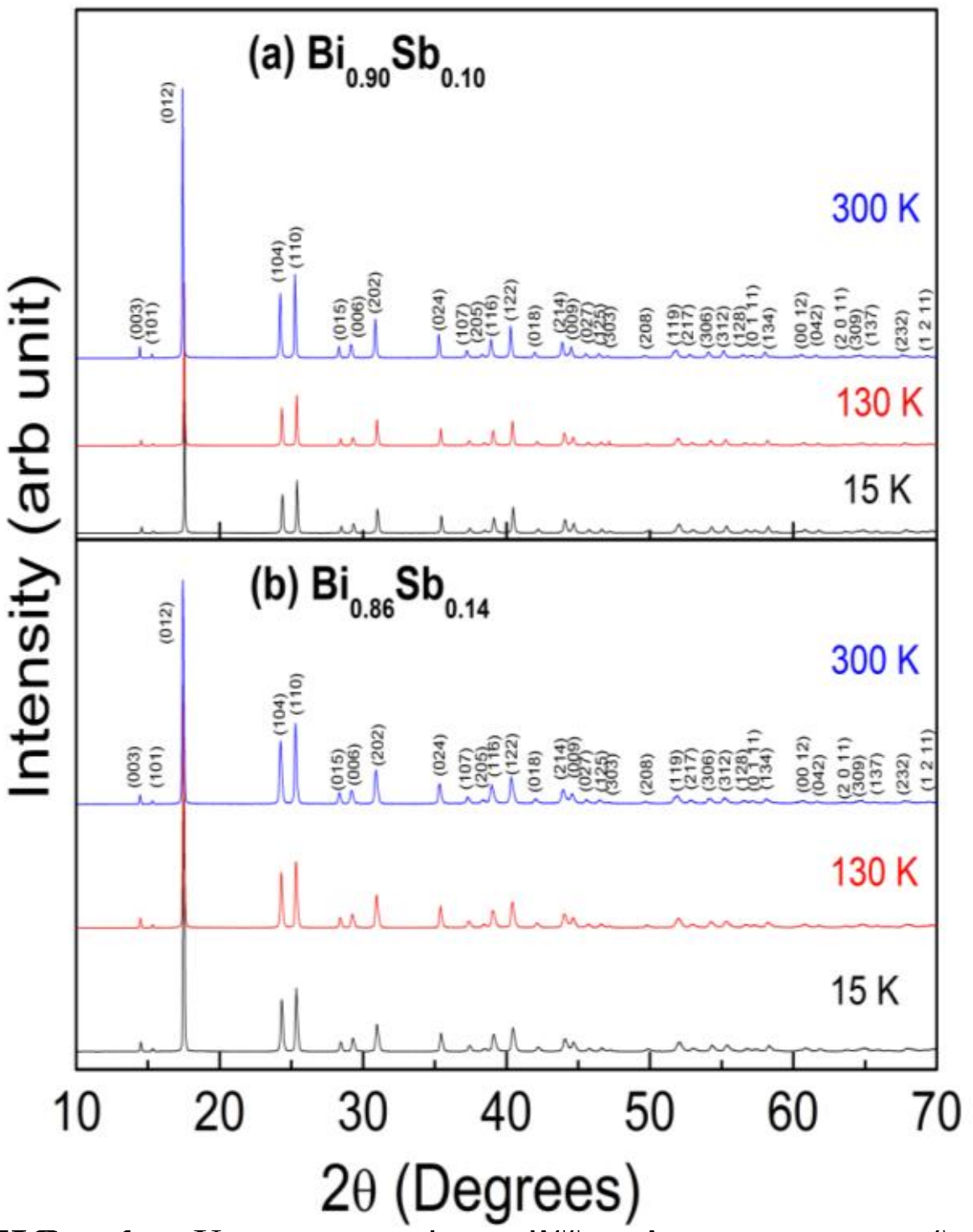

**FIG. 1.** X-ray powder diffraction patterns for $Bi_{0.90}Sb_{0.10}$ (a) and $Bi_{0.86}Sb_{0.14}$ (b) alloys at 300, 130 and 15 K.

The Debye-Waller factor ($B_{iso}$) that causes perturbation in band energies decreases significantly with increasing temperature [Table I].

The temperature dependence of resistivity for x=0.10 and 0.14 samples is shown in Fig. 2(a). For both the samples, ρ shows a non-monotonic temperature dependence. Temperature at which ρ displays a maximum has been designated as $T_{p\rho}$. Impurity and grain boundary scattering increase with increasing Sb concentration. This is reflected in the increased value of ρ in the reported $Bi_{1-x}Sb_x$ samples with increasing Sb content.[6] In addition, due to the presence of scattering sites in an alloy, *e-e* and *e-ph* scatterings significantly influence the transport property.[2,8] ρ, in low temperature regime, generally follows a $T^2$ dependence for the *e-e* interaction in metallic elements, alloys, and compounds.[15] Whereas the *e-ph* interaction contributes a $T^5$ dependence term at lower temperature ($T<<\theta_D$).[16] The Debye temperature ($\theta_D$) of binary alloys can be calculated from the Kopp-Neumann relation

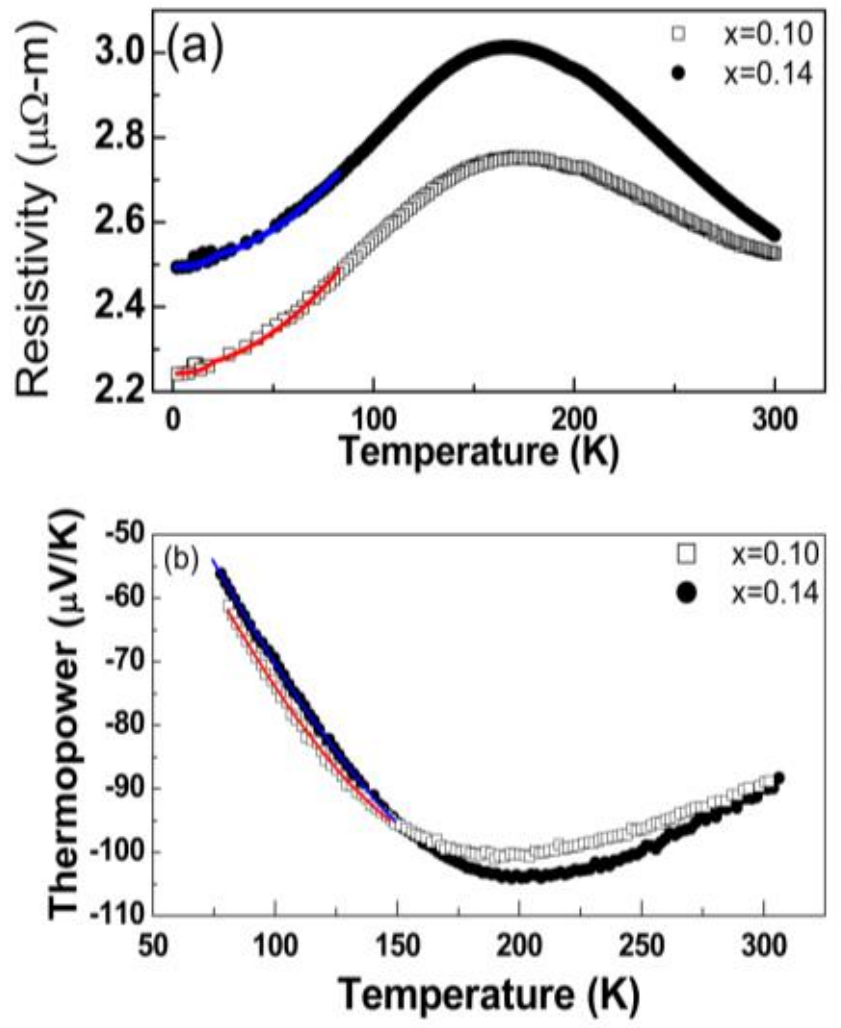

**FIG. 2.** Thermal variation of the electrical resistivity (a) and thermopower (b) for $Bi_{1-x}Sb_x$ alloys (x=0.10 and 0.14). Solid lines in (a) and (b) indicate the best fit with equations 2 and 3, respectively.

$$\theta_D^{-3} = (1-x)\theta_1^{-3} + x\theta_2^{-3}, \tag{1}$$

where $\theta_1$ and $\theta_2$ are the Debye temperatures of the solvent (Bi) and solute (Sb), respectively.[13] The estimated values of $\theta_D$ for the $Bi_{0.90}Sb_{0.10}$ and $Bi_{0.86}Sb_{0.14}$ samples are presented in Table II. In order to extract the contribution of the *e-e* and *e-ph* scattering on resistivity, the corresponding ρ(T) data below $\theta_D$ have been fitted with the equation

$$\rho = \rho_0 + A_1 T^2 + B_1 T^5, \tag{2}$$

where $\rho_0$ is a constant and coefficients $A_1$ and $B_1$ represent the *e-e* interaction and *e-ph* scattering strength, respectively. It is noteworthy to mention that, the *e-ph* scattering contribution is extracted for T $\leq$ 20K. The fitted values of $A_1$ and $B_1$ for both the samples are shown in Table II.

The temperature variation of thermoelectric power for the $Bi_{1-x}Sb_x$ alloys is shown in Fig. 2(b). Negative values of S throughout the temperature range of measurement imply that these samples are n-type in nature. With increasing temperature, the magnitude of S gradually increases and reaches a maximum value at $T_{pS}$. Beyond $T_{pS}$, S slowly decreases with the increase of temperature. Thermoelectric power of thin films and single crystals of $Bi_{1-x}Sb_x$ shows similar temperature dependence.[5,12] Thus, similar to resistivity, S(T) exhibits metallic-like behavior in the low-temperature regime below $T_{pS}$ and semiconducting-like behavior in the high-temperature region above $T_{pS}$. Absolute value of the Seebeck coefficient depends on Sb content. Careful observations reveal that in the semiconducting regime, the sample with higher $\rho$ possesses higher value of S. As in the case of resistivity, we have also estimated the contribution of *e-e* and *e-ph* scatterings on thermoelectric power. It should be mentioned that the $Bi_{1-x}Sb_x$ alloys behave as a degenerate conductor at low temperature with a small Fermi energy ($E_F$) due to the presence of impurity band.[5] The Seebeck coefficient in the low-temperature regime for such systems is represented by

$$S=A_2T + B_2T^3, \quad (3)$$

where $B_2$ ($=\pi^2k^2/3eE_F$) is the coefficient of the *e-ph* interaction term ($T^3$ dependency).[17] $A_2T$ is the contribution from diffusion thermopower, where electronic scattering plays a significant role.[17] The diffusion thermopower has a dominant contribution in the Seebeck coefficient of $Bi_{1-x}Sb_x$ and its value depends on the Sb concentration and the nature of the Fermi surface.[18] The $A_2$ and $B_2$ values obtained from the fitting of S(T) data using Eq. 3 are given in Table II. It has been shown that the band structure and carrier concentration of elemental Bi as well as $Bi_{1-x}Sb_x$ depend strongly on temperature.[18] $Bi_{1-x}Sb_x$ alloys in the high-temperature region behave like a nondegenerate intrinsic semiconductor. In such cases, the thermopower above $T_{pS}$ can be expressed as[12]

$$S = \frac{K_B}{e}\left(\frac{E_g}{2\mathrm{K}_B T} + B\right) \quad , \tag{4}$$

where $E_g$ and B are the band gap and scattering parameter, respectively. It has been assumed that the contributing bands are parabolic in nature with the same density of states and the carriers in this temperature regime are scattered primarily by the acoustic phonons. The best fit values of B, the parameter related to the scattering of acoustic phonons, are depicted in Table II. The estimated values of $E_g$ are also given in Table II. It is noteworthy to mention that the $E_g$ values calculated from S(T) data using Eq. 4 are consistent with those reported earlier from ρ(T) data.[6]

Power factor has been estimated from the ρ(T) and S(T) curves and plotted as a function of temperature in Fig. 3. Like ρ(T) and S(T), PF shows non-monotonic temperature dependence with its highest value around 200 K. Similar non-monotonic temperature dependence of PF has been reported earlier for single crystalline and thin film $Bi_{1-x}Sb_x$ systems with peak around 100 K and 250 K, respectively.[5,12] In polycrystalline samples, scatterings due to grain boundary and defects also play significant role. As the grain size for the x=0.14 sample is smaller than that for the x=0.10 sample, the scattering contribution due to these defects is expected to be larger for the former as compared to the later one.[6] This is consistent with the observed larger residual resistivity $\rho_0$ for the x=0.14 sample. Figure 3 clearly shows that the PF of the 10 at% Sb-doped sample is much higher. Further attempt has been made to determine the effect of *e-e* and *e-ph* scatterings on the estimated PF. We observe that the absolute values of the coefficients of *e-e* and *e-ph* scattering terms obtained by fitting both ρ(T) and S(T) curves in the low-temperature regime are higher for the $Bi_{0.90}Sb_{0.10}$ sample, which is a direct band gap material (Table II). This behavior is somehow unusual. In principle, indirect band gap material should have higher *e-ph* scattering. As $Bi_{1-x}Sb_x$ samples are narrow band gap semiconductor, the complex interplay of different bands, viz., $L_s$, $L_a$, H and T bands controls the transport property in these materials. This might lead to *e-ph* scattering enhancement in direct band gap $Bi_{0.90}Sb_{0.10}$ sample.

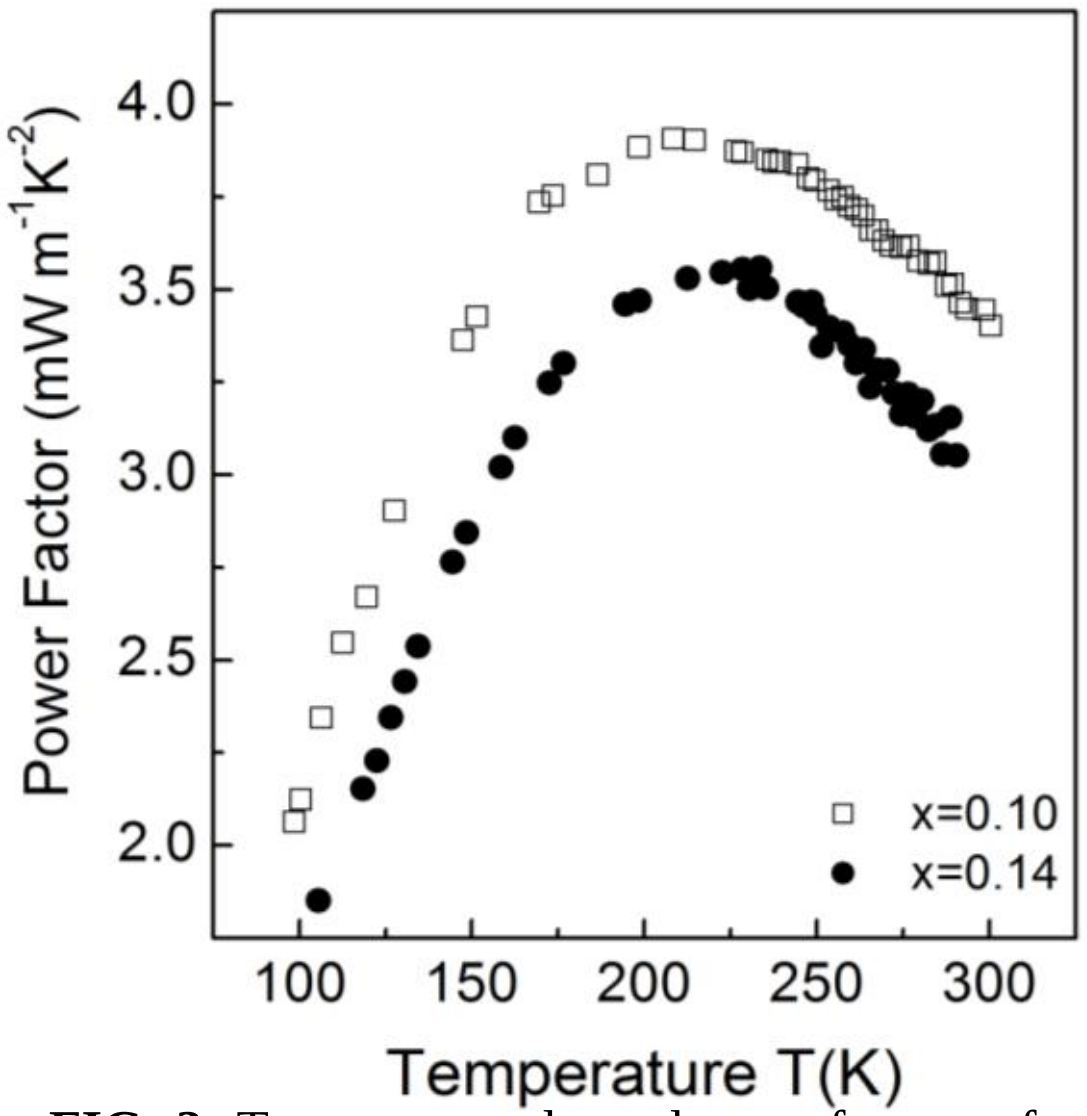

**FIG. 3.** Temperature dependence of power factor for $Bi_{1-x}Sb_x$ (x=0.10 and 0.14) alloys.

We would like to mention that the transport properties in the low-temperature region are dominated by both *e-e* and *e-ph* scatterings. However, the transport properties of $Bi_{1-x}Sb_x$ and related alloys in the high-temperature region are mainly controlled by the phonon scattering[4]. This insisted us in extracting the parameter related to the scattering of acoustic phonons (parameter B in Eq. 4) from the high-temperature regime of the S(T) data. The parameter B is also observed to be higher in $Bi_{0.90}Sb_{0.10}$. By maximizing the PF and lowering κ, ZT of a TE material can be improved. There is a long history of using atomic disorder to reduce lattice thermal conductivity. Phonon scattering plays a dominant role in an alloy because the random point defects created due to alloying act as ideal scattering sites.[2] It has been shown for skutterudite compounds like $CoSb_3$, carrier doping that optimizes ZT also substantially reduce κ through the *e-ph* interaction.[2] Recently, Dresselhaus *et al*. have demonstrated nanostructuring as an effective alternative route for enhancing ZT.[3] Actually, the phonon scattering at the interfaces between randomly oriented grains in nanostructured materials leads to a reduction in κ.[2,3] Extensive studies on SiGe alloys have shown that the polycrystalline materials with grain size around 1 μm have larger ZT as compared to single crystals. However, the evidence of enhanced phonon scattering as compared to electron scattering was demonstrated from experiments on SiGe materials with grains between 1-100 μm.[2] The grain size of $Bi_{1-x}Sb_x$ samples discussed in this work are comparable.[6] Hence, it is quite justified to assume that both *e-e* and *e-ph* scatterings concurrently affect the transport phenomenon viz., resistivity, thermopower and thermal conductivity but the phonon scattering should dominate as the temperature increases. Several efforts have been made to study the effect of scattering on κ.[2,3] But not much attention has been given to reveal the effect of scattering on PF. PF is also an important factor in controlling the efficiency of a TE material. Based on the results obtained from the analysis of resistivity and thermopower data of $Bi_{1-x}Sb_x$ (x=0.10, 0.14), we conclude that like thermal conductivity, PF is also sensitive to the *e-e* and *e-ph* scattering and the sample with the higher *e-e* and *e-ph* scattering terms possesses higher PF.

An attempt has been made to get an idea on the thermal expansion coefficient of $Bi_{1-x}Sb_x$ samples. Figure S3a (see the supplementary information for Figure S3a)[14] plots the measured lattice parameters for $Bi_{1-x}Sb_x$ (x=0.10, 0.14) alloys as a function of temperature, which have been fitted with a second-order polynomial. The linear thermal expansion coefficient of the material can be derived using the relation

$$\alpha = \frac{1}{a}\frac{da}{dT} \quad , \tag{5}$$

where $a$ is the lattice parameter. The temperature derivative of the fitted second-order polynomial has been utilized for estimating α using Eq. 5. The plots of α against temperature for both the samples are shown in Fig. S3b (see the supplementary information for Figure S3b).[14] Figure S3b indicates that temperature dependency of the estimated coefficient of linear expansion is stronger for the 10% Sb-doped sample. This is an important observation from the application point of view. Our observation based on resistivity and thermopower measurements depicts that the $Bi_{0.90}Sb_{0.10}$ sample possessing larger PF is a superior TE material as compared to the $Bi_{0.86}Sb_{0.14}$ sample. But the thermal expansion data predict that more care should be taken in designing a TE device with the $Bi_{0.90}Sb_{0.10}$ sample, as its coefficient of linear expansion is more sensitive to temperature. However, based on the lattice parameter values at three temperatures, estimation of linear expansion coefficient is a naïve approach. Precise calculation of linear expansion coefficient using more data points will give a better picture.[9,11]

In summary, in depth structural analyses have been done using the Rietveld refinement method. $Bi_{1-x}Sb_x$ samples retain the rhombhedral $A_7$ type structure down to the lowest temperature measured. The thermal expansion coefficient calculated from the temperature dependence of lattice parameter shows that thermal expansivity of the $Bi_{0.90}Sb_{0.10}$ sample depends strongly on temperature as compared to the $Bi_{0.86}Sb_{0.14}$ sample. Power factor is found to be higher for the $Bi_{0.90}Sb_{0.10}$ sample. The effect of the *e-e* and *e-ph* scatterings on resistivity and thermopower of $Bi_{1-x}Sb_x$ alloys have been estimated and it is concluded that the sample with higher *e-e* and *e-ph* scattering strength possesses higher PF.

This work is supported by the Department of Science and Technology (DST), Govt. of India and UGC, Govt. of India in the form of sanctioning research project, reference no. SR/FTP/PS-25/2009 and 39-990/2010(SR), respectively. KM is thankful to the University Grants Commission (UGC) for providing him Senior Research Fellowship and DD is grateful to DST for providing financial assistance in form of Junior Research fellowship through project no.

SR/FTP/PS-25/2009. The authors would also like to acknowledge DST, India and SINP, India for providing the experimental facilities at Indian Beamline, Photon Factory, Japan.

**Table I**. Structural parameters of hexagonal unit cell of $Bi_{0.90}Sb_{0.10}$ and $Bi_{0.86}Sb_{0.14}$ alloys at three different temperatures. The estimated values of $\chi^2$ and $B_{iso}$ are also shown.

| x | T(K) | *Hexagonal lattice parameter* $a_H$(Å) | $c_H$(Å) | V(Å$^3$) | $\chi^2$ | $B_{iso}$ Bi/Sb |
|---|---|---|---|---|---|---|
| 0.10 | 300 | 4.5359 $\pm$ 3.5 x $10^{-5}$ | 11.8361 $\pm$ 1.8 x $10^{-4}$ | 210.8949 | 1.472 | 0.041 |
| | 130 | 4.5291 $\pm$ 6.6 x $10^{-5}$ | 11.8099 $\pm$ 2.4 x $10^{-4}$ | 209.7977 | 1.559 | 1.074 |
| | 15 | 4.5262 $\pm$ 6.2 x $10^{-5}$ | 11.7967 $\pm$ 2.3 x $10^{-4}$ | 209.2949 | 1.519 | 1.338 |
| 0.14 | 300 | 4.5276 $\pm$ 1.2 x $10^{-4}$ | 11.8161 $\pm$ 3.5 x $10^{-4}$ | 209.7688 | 1.701 | 0.308 |
| | 130 | 4.5216 $\pm$ 7.8 x $10^{-5}$ | 11.7946 $\pm$ 3.1 x $10^{-4}$ | 208.8325 | 1.674 | 0.443 |
| | 15 | 4.5176 $\pm$ 1.3x $10^{-4}$ | 11.7766 $\pm$ 4.0 x $10^{-4}$ | 208.1451 | 1.590 | 0.829 |

**Table II**. The Debye temperature ($\theta_D$) and the best fit values of the parameters $A_1$, $B_1$, $A_2$, $B_2$, $E_g$, and B obtained from fitting the low temperature resistivity [Eq. 2], low temperature thermopower [Eq. 3] and high temperature thermopower [Eq. 4] data for $Bi_{0.90}Sb_{0.10}$ and $Bi_{0.86}Sb_{0.14}$ alloys.

| x | $\theta_D$ (K) | $A_1$ ($\Omega mK^{-2}$) | $B_1$ ($\Omega mK^{-5}$) | $\|A_2\|$ ($\mu VK^{-2}$) | $B_2$ ($\mu VK^{-4}$) | $E_g$ (meV) | B |
|---|---|---|---|---|---|---|---|
| 0.10 | 123.3 | 3.364 x $10^{-11}$ | 4.063 x $10^{-15}$ | 0.819 | 7.99 x $10^{-6}$ | 19.94 | 7.34 x $10^{-9}$ |
| 0.14 | 124.7 | 3.058 x $10^{-11}$ | 3.013 x $10^{-15}$ | 0.755 | 5.26 x $10^{-6}$ | 28.72 | 5.48 x $10^{-9}$ |

# Supplemental Material

**FIGURE S1:** Rietveld refinement pattern and the corresponding refinement parameters obtained using MAUD software for $Bi_{0.90}Sb_{0.10}$ alloy at (a) 300 K, (b) 130 K and (c) 15 K. The corresponding values of the estimated errors are also given.

(a)

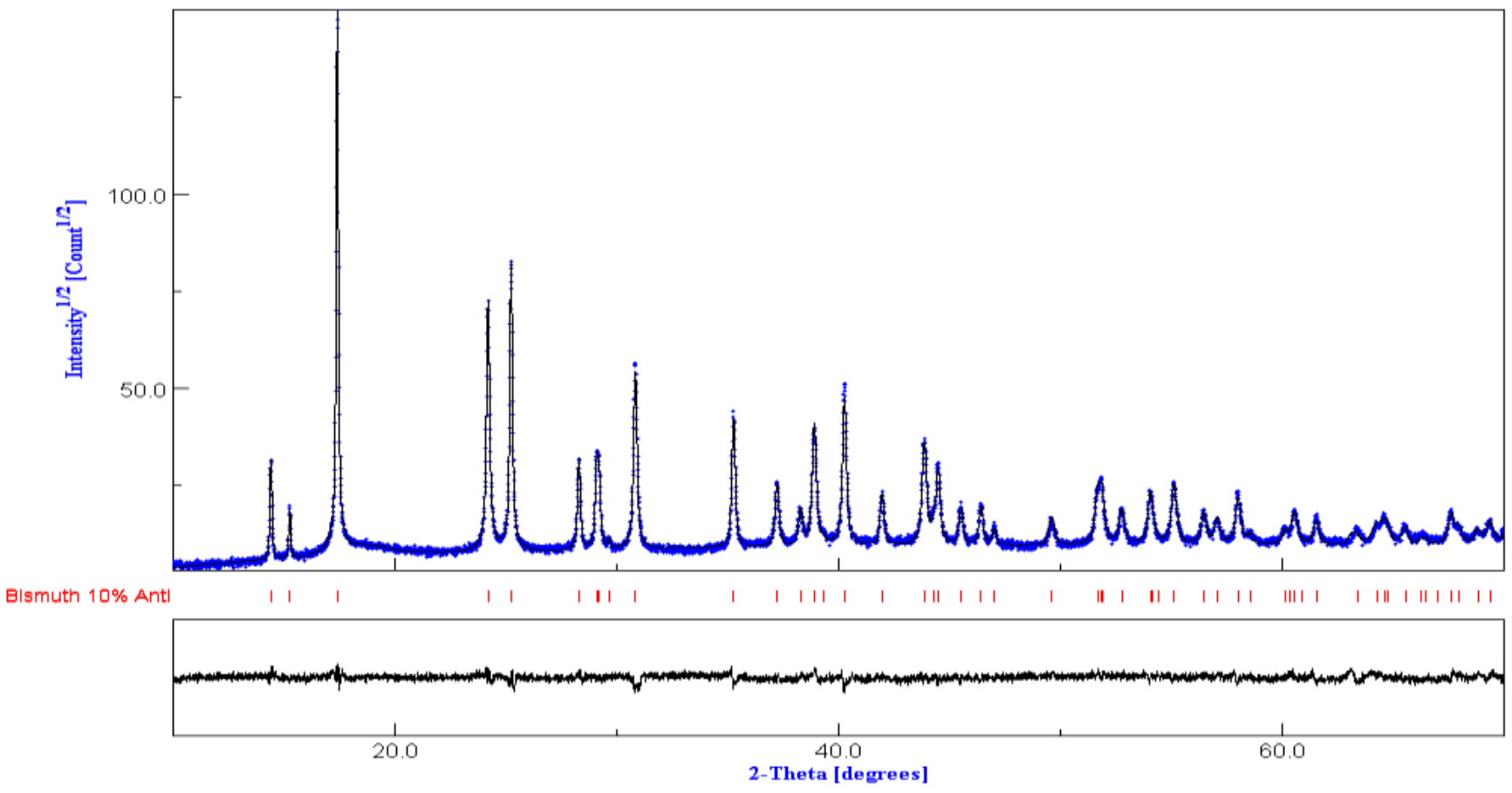

**300K**

| Phase | $Bi_{0.90}Sb_{0.10}$ [trigonal, $R\bar{3}m$ ] |
|---|---|
| Cell (Å) | *a*: 4.5359 (3.5 x $10^{-5}$); *c*: 11.8361 (1.8 x $10^{-4}$) |
| Size (nm)(hkl) | 123 (003), 210 (012), 96(104), 206(110), 142(202), 215(122), 206(300). |
| Strain(x$10^{-4}$) | 1.4 (003), 2.0 (012), 7.4(104), 3.6(110), 5.4(202), 3.6(122), 3.5 (300). |
| Bi $_{Occup.}$ | Bi1: 0.90 |
| Sb $_{Occup.}$ | Sb1: 0.10 |
| $Bi_x/Bi_y/Bi_z$ | 0.0 / 0.0/ 0.2341(4.3 x $10^{-5}$) |
| $Sb_x/Sb_y/Sb_z$ | 0.0 / 0.0/ 0. 2341 (equal Bi) |
| $B_{iso\ Bi/Sb}$ | Bi1: 0. 0405 (0. 018) Sb1: 0. 0405 (equal Bi) |
| $R_w$ (%) | 9.786 |
| $R_b$(%) | 7.481 |
| $R_{exp}$(%) | 6.650 |
| Gof | 1.472 |

(b)

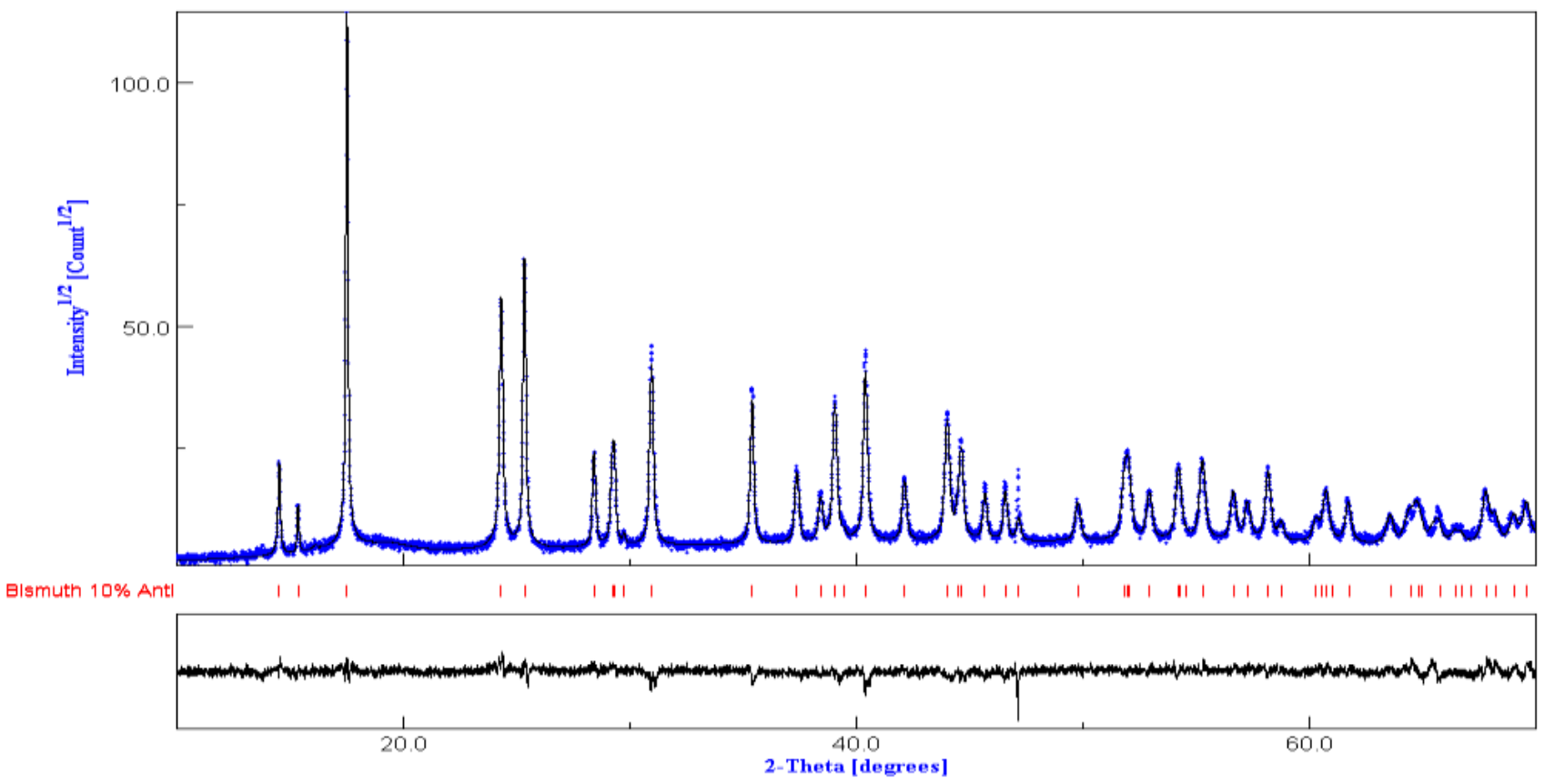

**130K**

| Phase | $Bi_{0.90}Sb_{0.10}$ [trigonal, $R\bar{3}m$ ] |
|---|---|
| Cell (Å) | *a*: 4.5291 (6.6 x $10^{-5}$); *c*: 11.8099 (2.4 x $10^{-4}$) |
| Size (nm)(hkl) | 72 (003), 124 (012), 74(104), 128(110), 103(202), 131(122), 128(300). |
| Strain(x$10^{-4}$) | 6.6 (003), 1.7 (012), 8.6(104), 6.1(110), 9.0(202), 4.7(122), 6.1 (300). |
| Bi $_{Occup.}$ | Bi1: 0.90 |
| Sb $_{Occup.}$ | Sb1: 0.10 |
| $Bi_x/Bi_y/Bi_z$ | 0.0 / 0.0/ 0.2343(5.0 x $10^{-5}$) |
| $Sb_x/Sb_y/Sb_z$ | 0.0 / 0.0/ 0. 2343 (equal Bi) |
| $B_{iso\ Bi/Sb}$ | Bi1: 1. 074 (0. 0183) Sb1: 1. 074 (equal Bi) |
| $R_w$ (%) | 13.634 |
| $R_b$(%) | 9.913 |
| $R_{exp}$(%) | 8.744 |
| Gof | 1.559 |

(c)

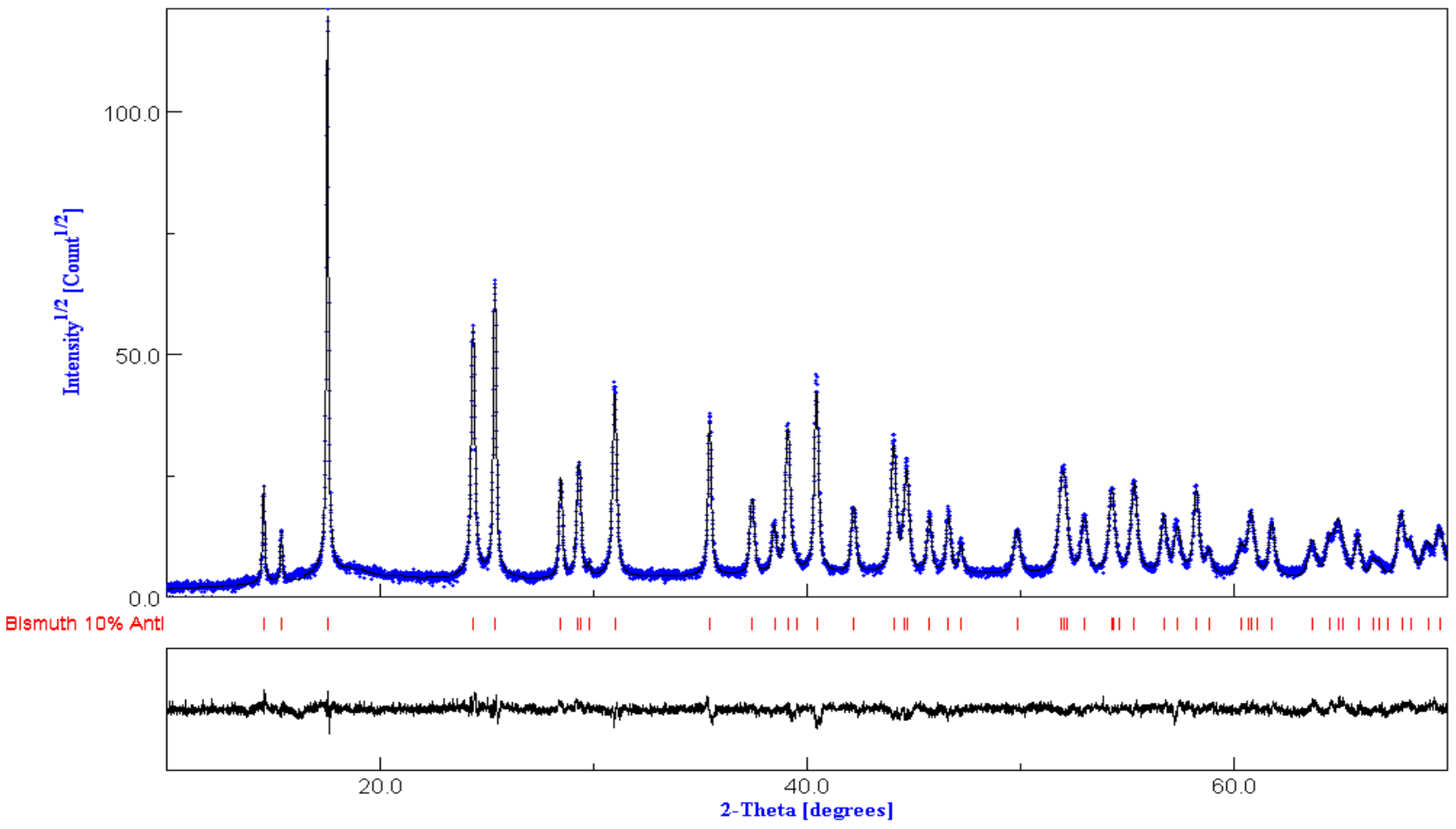

**15k**

| Phase | $Bi_{0.90}Sb_{0.10}$ [trigonal, $R\bar{3}m$ ] |
|---|---|
| Cell (Å) | *a*: 4.5262 (6.2 x $10^{-5}$); *c*: 11.7967 (2.3 x $10^{-4}$) |
| Size (nm)(hkl) | 79 (003), 137 (012), 73(104), 127(110), 93(202), 134(122), 127(300). |
| Strain(x$10^{-4}$) | 10.0 (003), 2.3 (012), 12.0(104), 7.3(110), 11.0(202), 5.4(122), 7.3 (300). |
| Bi $_{Occup.}$ | Bi1: 0.90 |
| Sb $_{Occup.}$ | Sb1: 0.10 |
| $Bi_x/Bi_y/Bi_z$ | 0.0 / 0.0/ 0.2342(4.2 x $10^{-5}$) |
| $Sb_x/Sb_y/Sb_z$ | 0.0 / 0.0/ 0. 2342 (equal Bi) |
| $B_{iso\ Bi/Sb}$ | Bi1: 1.338 (0. 016) Sb1: 1.338 (equal Bi) |
| $R_w$ (%) | 13.125 |
| $R_b$(%) | 9.447 |
| $R_{exp}$(%) | 8.642 |
| Gof | 1.519 |

**FIGURE S2:** Rietveld refinement pattern and the corresponding refinement parameters obtained using MAUD software for $Bi_{0.86}Sb_{0.14}$ alloy at (a) 300 K, (b) 130 K and (c) 15 K. The corresponding values of the estimated errors are also given.

(a)

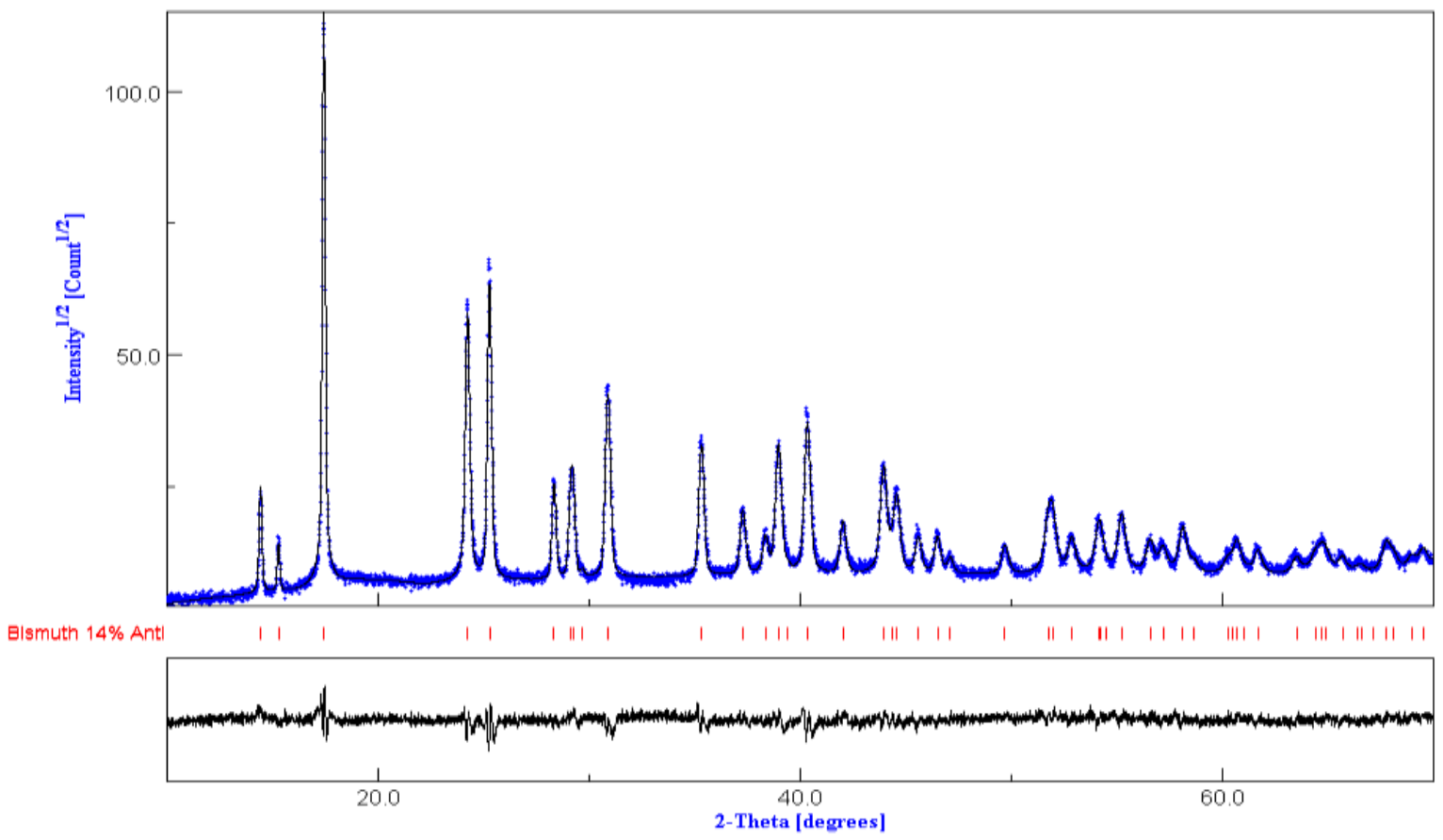

**300K**

| Phase | $Bi_{0.86}Sb_{014}$ [trigonal, $R\bar{3}m$ ] |
|---|---|
| Cell (Å) | *a*: 4.5276 ($1.2 \times 10^{-4}$); *c*: 11.8161 ($3.5 \times 10^{-4}$) |
| Size (nm)(hkl) | 71 (003), 111 (012), 56(104), 98(110), 67(202), 105(122), 98(300). |
| Strain($\times 10^{-4}$) | 0.5 (003), 1.3 (012), 0.7(104), 1.9(110), 1.8(202), 1.8(122), 1.9 (300). |
| Bi $_{Occup.}$ | Bi1: 0.86 |
| Sb $_{Occup.}$ | Sb1: 0.14 |
| $Bi_x/Bi_y/Bi_z$ | 0.0 / 0.0/ 0.2343($5.6 \times 10^{-5}$) |
| $Sb_x/Sb_y/Sb_z$ | 0.0 / 0.0/ 0.2343(equal Bi) |
| $B_{iso\ Bi/Sb}$ | Bi1: 0.308 (0.023) Sb1: 0.308 (equal Bi) |
| $R_w$ (%) | 12.256 |
| $R_b$(%) | 9.484 |
| $R_{exp}$(%) | 7.206 |
| Gof | 1.701 |

(b)

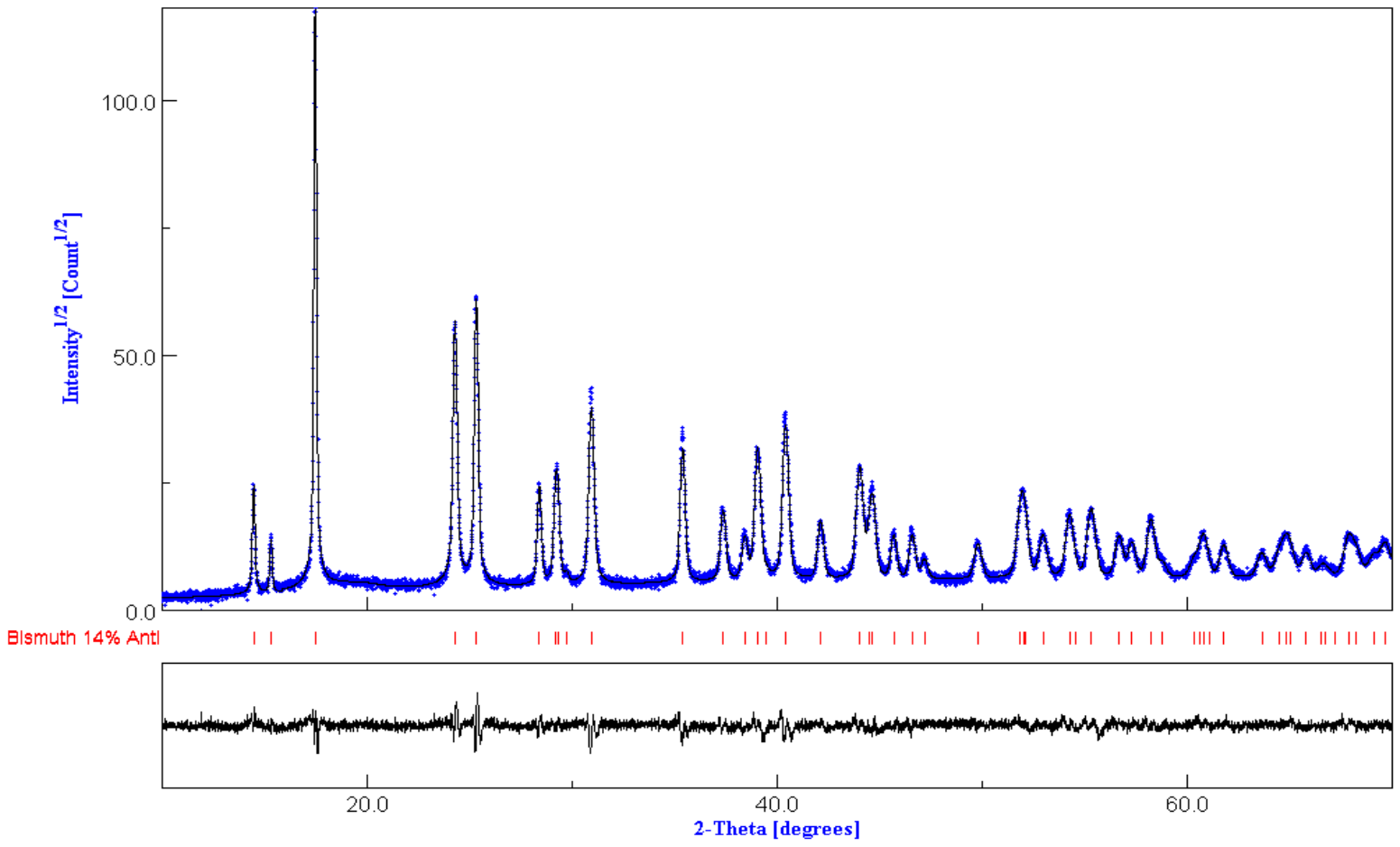

**130K**

| Phase | $Bi_{0.86}Sb_{014}$ [trigonal, $R\bar{3}m$ ] |
|---|---|
| Cell (Å) | *a*: 4.5216 ($7.8 \times 10^{-5}$); *c*: 11.7946 ($3.1 \times 10^{-4}$) |
| Size (nm)(hkl) | 71 (003), 130 (012), 75(104), 118(110), 92(202), 125(122), 118(300). |
| Strain($\times 10^{-3}$) | 2.0 (003), 2.1 (012), 2.3(104), 2.3(110), 2.5(202), 2.2(122), 2.3 (300). |
| Bi $_{Occup.}$ | Bi1: 0.86 |
| Sb $_{Occup.}$ | Sb1: 0.14 |
| $Bi_x/Bi_y/Bi_z$ | 0.0 / 0.0/ 0.2341($5.0 \times 10^{-5}$) |
| $Sb_x/Sb_y/Sb_z$ | 0.0 / 0.0/ 0.2341(equal Bi) |
| $B_{iso\ Bi/Sb}$ | Bi1: 0.443 (0.016) Sb1: 0.443 (equal Bi) |
| $R_w$ (%) | 13.120 |
| $R_b$(%) | 9.762 |
| $R_{exp}$(%) | 7.836 |
| Gof | 1.674 |

(c)

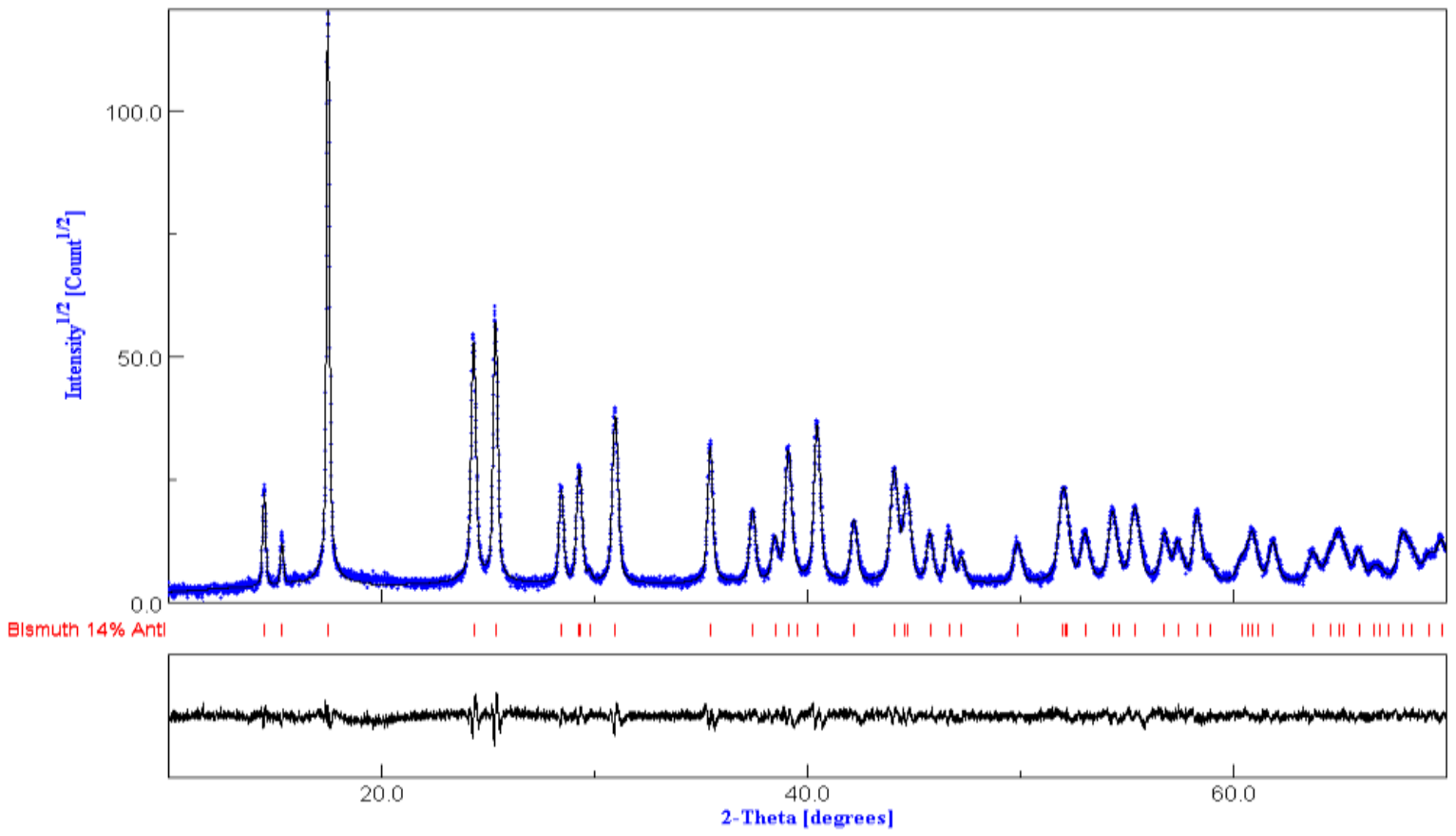

**15K**

| Phase | $Bi_{0.86}Sb_{014}$ [trigonal, $R\bar{3}m$ ] |
|---|---|
| Cell (Å) | *a*: 4.5176 ($1.3 \times 10^{-4}$); *c*: 11.7766 ($4.0 \times 10^{-4}$) |
| Size (nm)(hkl) | 95 (003), 73 (012), 66(104), 89(110), 76(202), 85(122), 89(300). |
| Strain($\times 10^{-3}$) | 1.6 (003), 0.4 (012), 1.7(104), 1.5(110), 1.9(202), 1.2(122), 1.5 (300). |
| Bi $_{Occup.}$ | Bi1: 0.86 |
| Sb $_{Occup.}$ | Sb1: 0.14 |
| $Bi_x/Bi_y/Bi_z$ | 0.0 / 0.0/ 0.2341($4.6 \times 10^{-5}$) |
| $Sb_x/Sb_y/Sb_z$ | 0.0 / 0.0/ 0.2341(equal Bi) |
| $B_{iso\ Bi/Sb}$ | Bi1: 0.829 (0.019) Sb1: 0.829 (equal Bi) |
| $R_w$ (%) | 13.177 |
| $R_b$(%) | 9.339 |
| $R_{exp}$(%) | 8.285 |
| Gof | 1.590 |

**FIGURE S3a:** Lattice constant of $Bi_{1-x}Sb_x$ (x=0.10, 0.14) as a function of temperature. The solid lines represent the best fit second order polynomial.

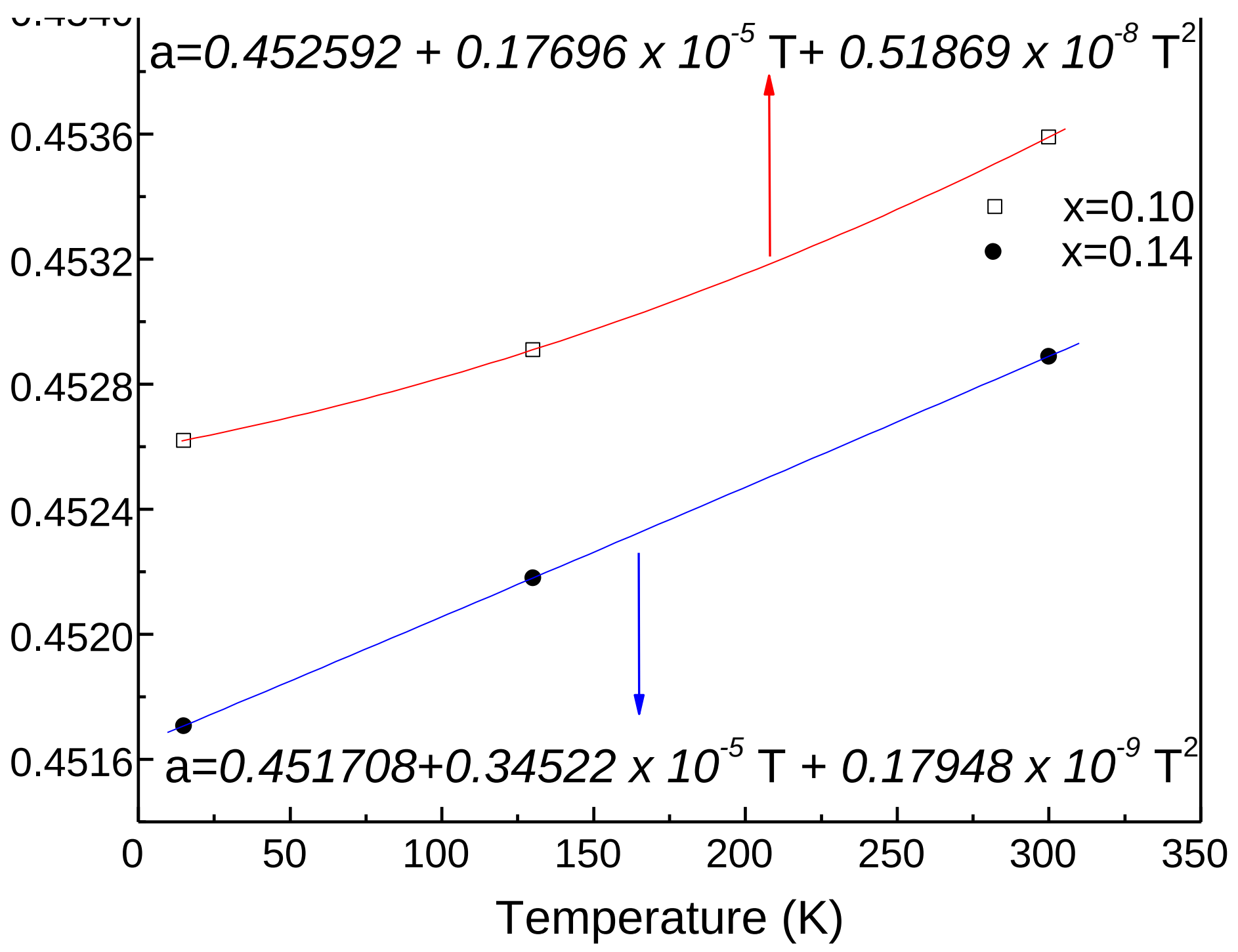

**FIGURE S3b:** Linear thermal expansion coefficients of $Bi_{0.90}Sb_{0.10}$ and $Bi_{0.86}Sb_{0.14}$ as a function of temperature. Solid line is a guide to the eye.

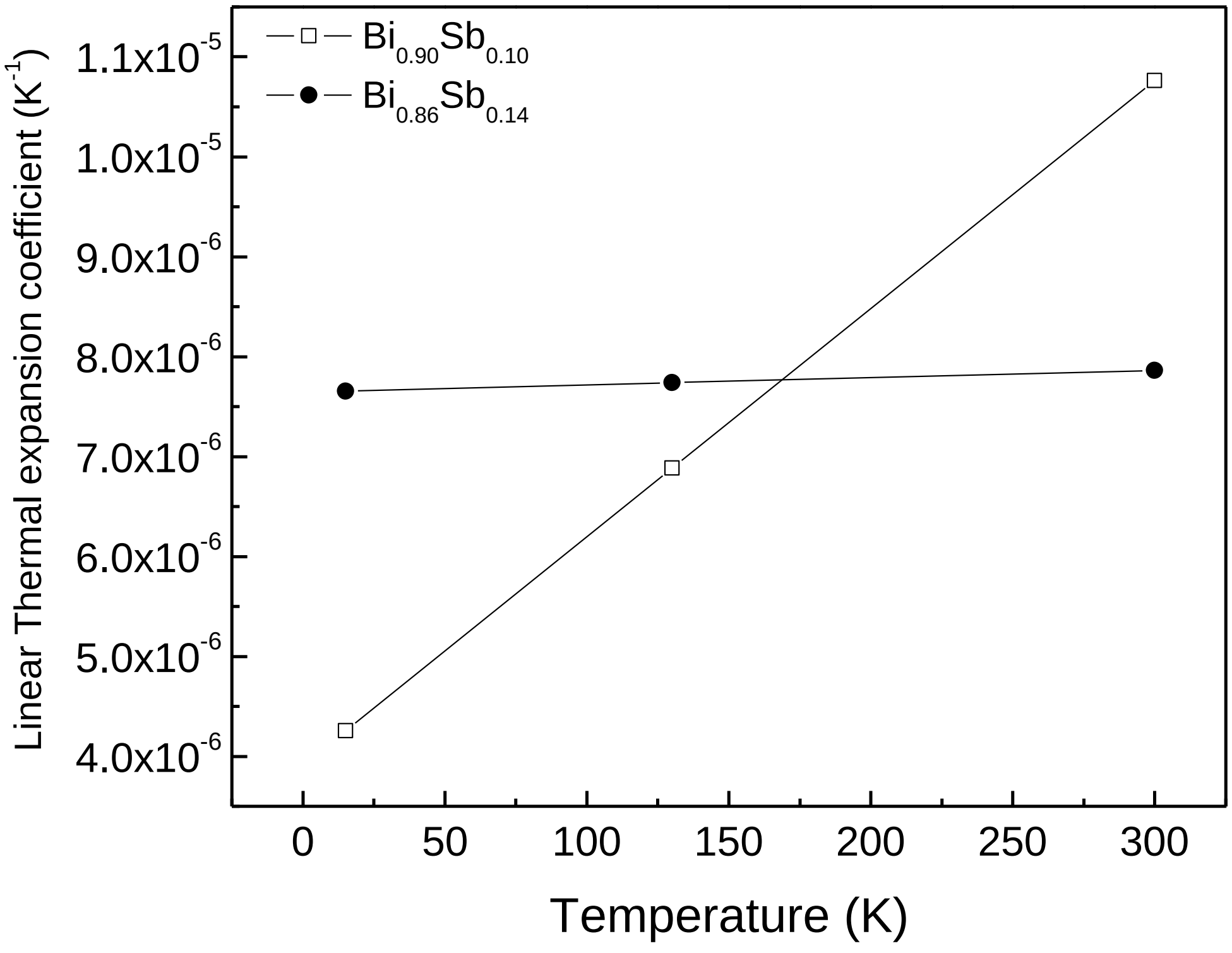